 \documentclass[aps, twocolumn, showpacs, amsmath]{revtex4}
\usepackage{dcolumn}
\usepackage{bm}
\usepackage{graphicx}
\usepackage{color}
\usepackage{subfigure}
\usepackage{hyperref}
\usepackage{latexsym}
\usepackage{amsthm}
\usepackage{amssymb}

\DeclareGraphicsExtensions{.jpg,.pdf, .mps, .png, .eps, .ps, .EPS,.gif}
\begin{document}
\def\be{\begin{equation}}
\def\ee{\end{equation}}

\def\bc{\begin{center}}
\def\ec{\end{center}}
\def\bea{\begin{eqnarray}}
\def\eea{\end{eqnarray}}
\newcommand{\avg}[1]{\langle{#1}\rangle}
\newcommand{\Avg}[1]{\left\langle{#1}\right\rangle}

\def\ie{\textit{i.e.}}
\def\etal{\textit{et al.}}
\def\m{\vec{m}}
\def\G{\mathcal{G}}

\title{Fluctuations in percolation  of  sparse complex networks}

\author{ Ginestra Bianconi}
\affiliation{School of Mathematical Sciences, Queen Mary University of London, London, E1 4NS, United Kingdom}

\begin{abstract}
We study the role of fluctuations in percolation of sparse complex networks. To this end we consider two random correlated realizations of the initial damage of the nodes and we evaluate the fraction of nodes that are expected to remain in the giant component of the network  in both cases or just in one case. Our framework includes a message-passing algorithm able to predict the fluctuations in a single network, and an analytic  prediction of the expected fluctuations in  ensembles of sparse networks. This approach is applied to real ecological and infrastructure networks and it is shown to characterize the expected fluctuations in their response to external damage. 
\end{abstract}


\maketitle
\section{Introduction}
 Percolation  is one of the most interesting and fundamental critical  phenomena \cite{crit,Alain} defined on complex networks \cite{NS,Newman_book}. It  characterizes the non-linear response of a network to random damage of its nodes (or links) by evaluating the size of the giant component that results after the initial perturbation.  In network science percolation  has received  ever-lasting attention and currently  methods and ideas developed in the framework of  percolation theory are widely used to study social, technological and biological networks. At the beginning of field,  percolation theory on  complex networks has been pivotal to characterize the  robustness of scale-free networks \cite{Cohen1,Cohen2,Newman_old1,Newman_old2,Laszlo_robustness}. More recently generalized percolation processes including k-core percolation \cite{Doro_k_core}, bootstrap  percolation \cite{Doro_bootstrap} and percolation of  multilayer networks \cite{Havlin1,Havlin2,Stanley,Raissa,Son,Baxter,BD2,Goh,Baxter2016,Redundant,Antagonist,Bond} have greatly enriched our understanding of the interplay between the structure of  networks and their response to perturbations. 

In locally tree-like networks,  percolation  can be studied using message passing algorithms \cite{Mezard,Weigt}.  These algorithms are becoming increasingly popular in network theory and they have been used to characterize the percolation of single \cite{Lenka} and  multilayer networks \cite{Cellai2013,Cellai2016,Kabashima,Redundant,Radicchi,BiRa}, to predict and monitor epidemic spreading \cite{Newman_epidemics,Dallasta1,Dallasta2,Bianconi_Epidemics,Saad2,Gleeson}, to identify the driver nodes of a network ensuring its controllability \cite{Control,Bianconi_control1} and to solve a number of other optimization problems on networks \cite{Saad1,Makse,Dismantling}. 

In this paper we aim at using a message-passing algorithm valid in the locally tree-like approximation, to evaluate  the fluctuations that can be observed in the response of a network to random damage. This problem is of wide interest for the network science community and can be applied to a variety of real  biological, social and technological networks to gain a comprehensive understanding of their robustness properties.

In all percolation-like studies the goal is to characterize the fraction of nodes in the giant component (or in the considered generalization of the giant component) after an initial damage is inflicted to the nodes (or the links) of the network. However, it is usually the case that the real entity of the initial damage is not known. Instead often only  the probability that a random node or a random link of the network is initially damaged is known.
In this case it is standard  to characterize the response of the network to the external perturbation by considering the expected fraction of nodes remaining in the giant component (or in its generalization) after a random  initial damage.
For instance, very reliable predictions of this average response of a single network to a random damage of its nodes or links can be obtained by message-passing techniques \cite{Lenka} as long as the network is locally tree-like. 
Our aim here is to go beyond this approach  proposing a framework able to characterize the fluctuations observed in the response of a network to different realizations of the initial damage considering  also the case in which these initial perturbations are correlated.
To start with a simple case, we address exclusively percolation of single sparse networks  (i.e. the emergence of the giant component).
Given  two random realizations of the initial damage, where the second realization of the initial damage can be correlated with the first realization of the initial damage,  we  characterize which is the probability that a node is found in the giant component in both realizations or just in one realization of the initial damage. In this way we  identify when the network has the most unpredictable response to damage. This point is  signalled by a maximum in the fraction of nodes that are found in the giant component for one realization of the damage but are not found in the giant component for the other realization of the damage.
The proposed   message-passing algorithm is here tested over real networks including food-webs and infrastructure networks.
Finally  the critical behavior observed in uncorrelated sparse network ensembles with given degree distribution is here characterized by deriving the relevant  critical indices.
\section{The message passing algorithm}
\subsection{The message passing algorithm for single realizations of the initial damage} Consider two different realizations of the initial damage of the nodes indicated respectively by  $q=1,2$. Each realization of the initial damage $q=1,2$, is fully characterized by the set of variables  $\{s_i^{(q)}\}_{i=1,2,\ldots, N}$ where  $s_i^{(q)}$ indicates whether a node $i$ is initially removed ($s_i^{(q)}=0$) or not ($s_i^{(q)}=1$) from the network.
In a locally tree-like network, a well known message passing algorithm \cite{Mezard,Weigt} is able to predict   whether a node $i$ belongs ($n_i^{(q)}=1$) or not ($n_i^{(q)}=0$) to the giant component after the initial damage indicated by $\{s_i^{(q)}\}_{i=1,2,\ldots, N}$ has been inflicted to the network.
Specifically the values of the indicator functions $n_i^{(q)}$ are determined by a set of messages $n_{i\to j}^{(q)}$ that are exchanged between connected nodes $i$ and $j$.
These messages take values zero or one (i.e. $n_{i\to j}^{(q)}=0,1$) and indicate whether $(n_{i\to j}^{(q)}=1)$ or not ($n_{i\to j}^{(q)}=0$) node $i$ connects node $j$ to other nodes in the giant component. These messages are  determined by the following recursive set of equations
\bea
n_{i\to j}^{(q)}=s_i^{(q)}\left(1-\prod_{\ell\in N(i)\setminus j}(1-n_{\ell\to i}^{(q)})\right),
\eea
where $N(i)$ indicates the set of neighbors of node $i$.
In other words node $i$  connects node $j$ to nodes in the giant component ($n_{i\to j}^{(q)}=1$) if and only if it is not initially damaged (i.e. $s_i^{(q)}=1$) and it has at least a neighbor node $\ell$ different from node $j$ that at its turn connects node $i$ to other nodes in the giant component. 
The messages  $n_{i\to j}^{(q)}$ determine the value of the indicator functions $n_i^{(q)}$. Each indicator function  is set equal to one (i.e. $n_i^{(q)}=1$)  if and only if node $i$ is not initially damaged (i.e. $s_i^{(q)}=1$) and it has at least a neighbor node $\ell$ that connects it to other nodes in the giant component, (i.e. $n_{\ell\to i}^{(q)}=1$).
Therefore we have that the indicator functions $n_i^{(q)}$ are determined by 
\bea
n_i^{(q)}=s_i^{(q)}\left(1-\prod_{\ell\in N(i)}(1-n_{\ell\to i}^{(q)})\right).
\eea

\subsection{Message passing algorithm to evaluate fluctuations}
 It is often the case that the exact realization of the initial random damage is not known, and only the probability that the initial damage occurs on any given node of the network is available. 
It order to treat this scenario, the probability that a node is in the giant component is usually studied \cite{Lenka}. When two independent realizations of the initial damage are applied to a given network the response  show fluctuations. These fluctuations can become highly non-trivial in the case in which the two realizations of the initial damage are correlated.
To  characterize the fluctuations  in the general case of node-dependent and correlated damage, we consider two realizations ($q=1,2$) of the initial  random damage.  Each node $i$ is  damaged just in one or in both realizations of the damage with a node-dependent probability. It follows that  in a pair $q=1,2$ of realizations of the initial damage,  the initial damage configuration $\{s_i^{(1)},s_i^{(2)}\}_{i=1,2,\ldots, N}$ has probability 
\bea
\hat{P}(\{s_i^{(1)},s_i^{(2)}\})&=&\prod_{i=1}^N \left[\left(p_i^{[11]}\right)^{s_i^{(1)}s_i^{(2)}}\left(p_i^{[00]}\right)^{(1-s_i^{(1)})(1-s_i^{(2)})}\right.\nonumber \\&&\left.\left(p_i^{[10]}\right)^{s_i^{(1)}(1-s_i^{(2)})}\left(p_i^{[01]}\right)^{(1-s_i^{(1)})s_i^{(2)}}\right],
\label{hP}
\eea
where $p_i^{[11]},p_i^{[01]},p_i^{[10]}$ and $p_i^{[00]}$ indicate respectively the probability that node $i$ is not initially damaged for both $q=1$ and $q=2$; the probability that it is initially damaged for $q=1$ and not for $q=2$; the probability that it  is not initially damaged for $q=1$ and is initially damaged for $q=2$; or the probability that  it is initially damaged for both $q=1$ and $q=2$.
Note that for every node $i=1,2,\ldots, N$ these probabilities are normalized, and  we have  
\bea
p_i^{[11]}+p_i^{[01]}+p_i^{[10]}+p_i^{[00]}=1.
\label{norm}
\eea
Here and in  the following we will  indicate with $p_i^{(q)}$ the probability that a node $i$ is not  initially damaged in the realization $q$, these probabilities are given by 
\bea
p^{(1)}_i&=&p_i^{[10]}+p_i^{[11]},\nonumber \\
p^{(2)}_i&=&p_i^{[01]}+p_i^{[11]}.
\eea
When we consider two configurations of the initial damage drawn from the distribution $P(\{s_i^{(1)},s_i^{(2)}\})$ given by Eq. $(\ref{hP})$ the probability $\sigma_i^{(q)}$ that a node $i$ is in the giant component of the network in the $q$-th realization of the initial damage  is given by   
\bea{\sigma}_i^{(q)}&=&\Avg{n_i^{(q)}}
\eea
 where $\Avg{\ldots}$ indicates the average  over the  probability distribution $\hat{P}(\{s_i^{(1)},s_i^{(2)}\})$.
In order to go beyond this description here we  study the probability  that node $i$ is  in the giant component for both  realizations of the initial random  damage ($\hat{\sigma}^{[11]}_i$),  the probability that node $i$ is in the giant component only for the first realization of the random damage ($\hat{\sigma}^{[10]}_i$), the probability that it is in the giant component only for the second realization of the random damage ($\hat{\sigma}^{[01]}_i$), and finally the probability that it is not in the giant component for both realizations of the random damage ($\hat{\sigma}^{[00]}_i$). These probabilities are given by 
\bea
\hat{\sigma}_i^{[11]}&=&\Avg{n_i^{(1)}n_i^{(2)}},\nonumber \\
\hat{\sigma}_i^{[10]}&=&\Avg{n_i^{(1)}\left(1-n_i^{(2)}\right)}=\sigma_i^{(1)}-\hat{\sigma}_i^{[11]},\nonumber \\
\hat{\sigma}_i^{[01]}&=&\Avg{\left(1-n_i^{(1)}\right)n_i^{(2)}}=\sigma_i^{(2)}-\hat{\sigma}_i^{[11]},\nonumber \\
\hat{\sigma}_i^{[00]}&=&\Avg{\left(1-n_i^{(1)}\right)\left(1-n_i^{(2)}\right)}\nonumber \\
&=&1-\sigma_i^{(1)}-\sigma_i^{(2)}+\hat{\sigma}_i^{[11]}.
\label{sigma}
\eea
 where here $\Avg{\ldots}$ indicates the average  over the  probability distribution $\hat{P}(\{s_i^{(1)},s_i^{(2)}\})$.
 From Eqs. $(\ref{sigma})$ it is evident that given $\sigma_{i}^{(1)}, \sigma_i^{(2)}$ and $\hat{\sigma}_i^{[11]}$ all the remaining probabilities  can be calculated.
 In order to evaluate $\sigma_i^{(q)}$ and $\hat{\sigma}_i^{[11]}$ we need to find the average messages ${\sigma}_{i\to j}^{(q)}=\Avg{n_{i\to j}^{(q)}}$ and 
$\hat{\sigma}_{i\to j}^{[11]}=\Avg{n_{i\to j}^{(1)}n_{i\to j}^{(2)}}$ over the distributions $P(\{s_i^{(q)}\})$.
The equations determining  the indicator functions $\sigma_i^{(1)},\sigma_i^{(2)}$ and $\hat{\sigma}_i^{[11]}$ and the corresponding messages are  given, in a  locally tree-like network, on one side by the well known message passing equations \cite{Lenka}
	\bea	
	\hat{\sigma}_{i\to j}^{(q)} &=&p^{(q)}_i\left[1- \prod_{\ell \in N(i)\setminus j}\left(1-  {\sigma}_{\ell \to i}^{(q)} \right)\right],\nonumber \\	
	\hat{\sigma}_{i}^{(q)} &=&p^{(q)}_i\left[1-\prod_{\ell \in N(i)}\left(1-  {\sigma}_{\ell \to i}^{(q)}\right)\right],
	\label{MP_eq1}
	\eea
	for $q=1,2$ and on the other side by the additional set of equations, introduced here to account for fluctuations, 
	\bea
	\hat{\sigma}_{i\to j}^{[11]} &=&p^{[11]}_i\left[1- 	\prod_{\ell \in N(i)\setminus j}\left(1-  {\sigma}_{\ell \to i}^{(1)} \right)\right.\nonumber \\
	&&- \prod_{\ell \in N(i)\setminus j}\left(1-  {\sigma}_{\ell \to i}^{(2)} \right)\nonumber \\
	&&+\left.\prod_{\ell \in N(i)\setminus j}\left(1-  {\sigma}_{\ell \to i}^{(1)}- {\sigma}_{\ell \to i}^{(2)} +\hat{\sigma}_{\ell \to i}^{[11]}\right)\right],\nonumber \\
	\hat{\sigma}_{i}^{[11]} &=&p^{[11]}_i\left[1- \prod_{\ell \in N(i)}\left(1-  {\sigma}_{\ell \to i}^{(1)} \right)\right.\nonumber \\
	&&- \prod_{\ell \in N(i)}\left(1-  {\sigma}_{\ell \to i}^{(2)} \right)\nonumber \\
	&&+\left.\prod_{\ell \in N(i)}\left(1-  {\sigma}_{\ell \to i}^{(1)}- {\sigma}_{\ell \to i}^{(2)} +\hat{\sigma}_{\ell \to i}^{[11]}\right)\right].
	\label{MP_eq2}
\eea
Note that now both messages and indicator functions take real values between zero and one.

	\begin{figure*}
    \includegraphics[width=1.8\columnwidth]{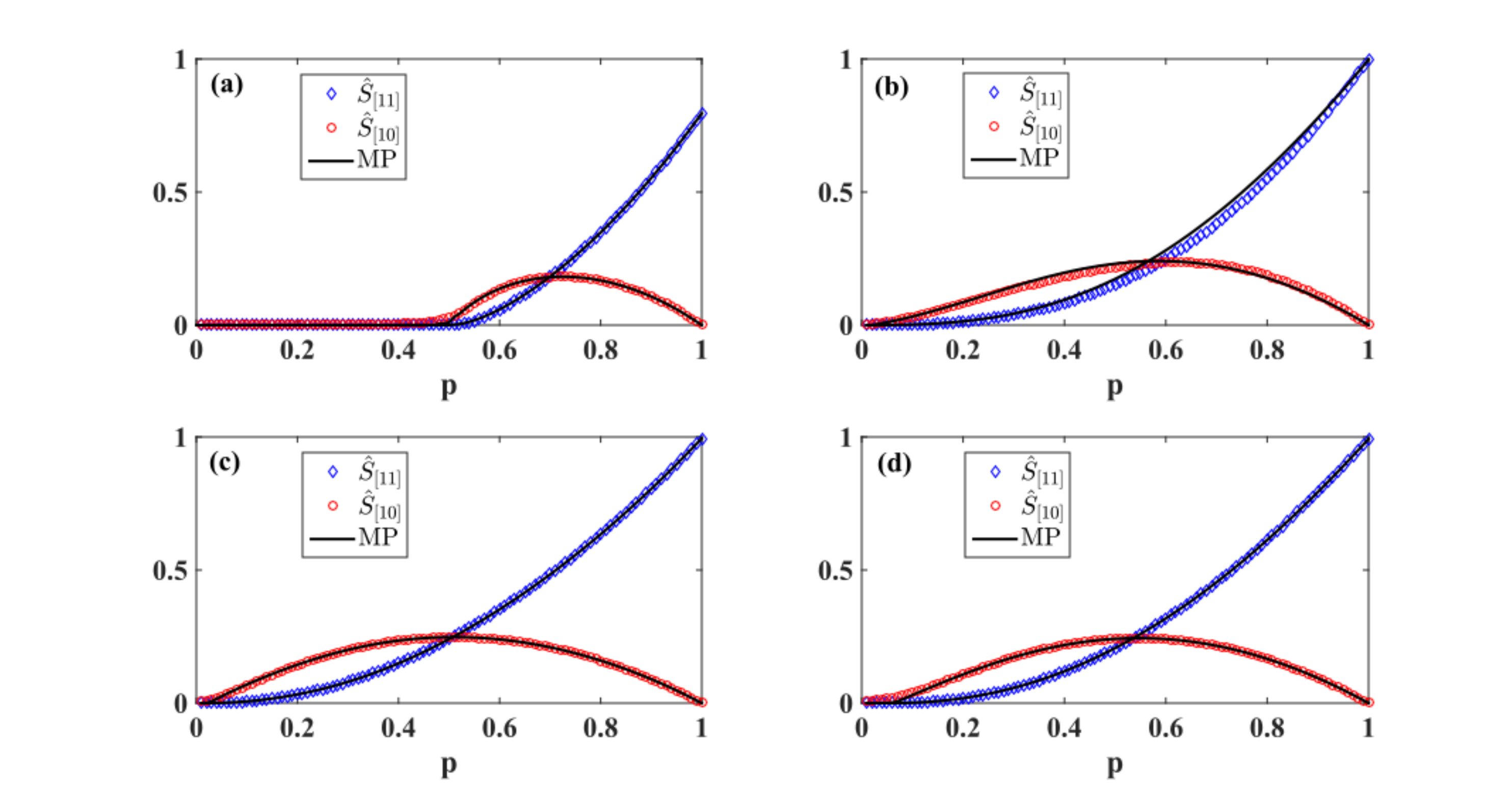}
	\caption{(Color online) The fraction of nodes  $\hat{S}_{[11]}$ and  $\hat{S}_{[10]}$  which are respectively in the giant component in two random realizations of the initial damage and just in one realization of the initial damage are plotted as a function of the probability that a node is not initially damaged in any realization of the initial damage $p^{(1)}_i=p^{(2)}_i=p$. Here only the case  in which the  two  realizations of the initial damage are uncorrelated $p^{[11]}_i=p^2$, for $i=1,2,\ldots, N$ is considered. The reported simulation results are compared with the message passing predictions (MP) on the same single network for four networks: (a) a random Poisson network with average degree $\avg{k}=2$ and total number of nodes $N=10^4$; (b) the US airport network \cite{USairports}; (c) the Little Rock Lake Foodweb Network \cite{Littlerock,data}; (d) The Ythan Estuary Foodweb Network \cite{data}. The simulation results are obtained by averaging over $5000$ pairs of random realization of the initial damage.}
	\label{fig:MP_simulations}
\end{figure*}
In the case of uncorrelated initial damage 
 when for every node $i$ we have 
 \bea
 p^{[11]}_i=p_i^{(1)}p_i^{(2)}
 \eea
 the Eqs. $(\ref{MP_eq2})$ have always the trivial solution 
\bea
\sigma_i^{[11]}&=&\sigma_i^{(1)}\sigma^{(2)}_i,\nonumber \\
\sigma_{i\to j}^{[11]}&=&\sigma^{(1)}_{i\to j}\sigma^{(2)}_{i\to j}.
\eea
Additionally in the  case in which $p^{(1)}_i=p^{(2)}_i$ for every node $i$ the Eqs. $(\ref{MP_eq1})$ simplify since we have  
\bea
\sigma_i^{(1)}&=&\sigma^{(2)}_i,\nonumber \\
\sigma_{i\to j}^{(1)}&=&\sigma^{(2)}_{i\to j}.
\eea

In order to characterize the global response of the network to the initial damage  it is convenient to consider  the expected fraction $\hat{S}_{[11]}$ of nodes that are in the giant component in both realization of the initial damage, and the expected fraction $\hat{S}_{[10]}$, ($\hat{S}_{[01]}$) of nodes that are in the giant component just in the first (second) realization of the initial damage. These are clearly given by 
\bea
\hat{S}_{[r,r']}=\frac{1}{N}\sum_{i=1}^N \hat{\sigma}_i^{[rr']},
\eea
where here and in the following  $[rr']$ can take values $[11],[10]$ or $[01]$.

We note here that strictly speaking $\hat{S}_{[10]}$ characterizes the fluctuations in the response to initial damage only if the two realizations of the initial damage are statistically equivalent, i.e. for $p_i^{(1)}=p_i^{(2)}$ while for
$p_i^{(1)} \neq p_i^{(2)}$ the use of the term  fluctuations is less appropriate.

\subsection{Numerical results on single networks}
\begin{figure*}
    \includegraphics[width=1.8\columnwidth]{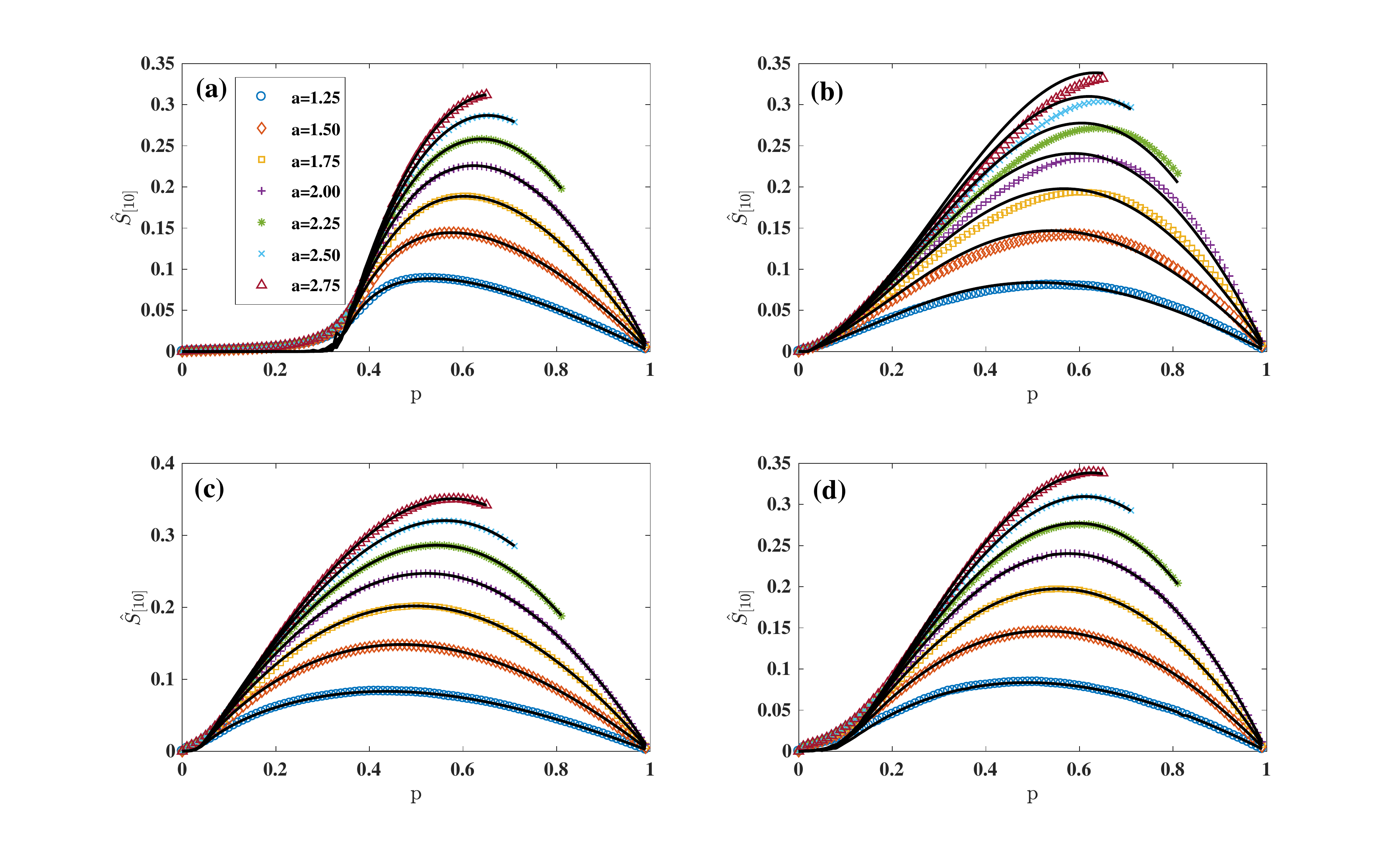}
	\caption{(Color online) The fraction of nodes    $\hat{S}_{[10]}$  which are in the giant component in  just in one realization of the initial damage is plotted as a function of the probability that a node $i$ is not initially damaged in any realization of the initial damage $p^{(1)}_i=p^{(2)}_i=p$. The data are shown for $p^{[11]}_i=p^a$ and $a=1.25, 1.50, 1.75$ (positively correlated case), $a=2.00$ (uncorrelated case) and $a=2.25,2.50,2.75$ (negatively correlated case). The reported simulation results (symbols) are compared with the message passing predictions (solid lines) on the same single network for four networks: (a) a random Poisson network with average degree $\avg{k}=3$ and total number of nodes $N=2\times 10^3$; (b) the US airport network \cite{USairports}; (c) the Little Rock Lake Foodweb Network \cite{Littlerock,data}; (d) The Ythan Estuary Foodweb Network \cite{data}. The simulation results are obtained by averaging over $5000$ pairs of random realization of the initial damage.}
	\label{fig:MP_simulations2}
\end{figure*}

 In order to validate our theoretical description of fluctuations in the percolation properties of single networks, we have compared the results obtained by applying the message passing algorithm described by Eqs. $(\ref{MP_eq1})-(\ref{MP_eq2})$ to simulations of random damage on single networks (the code is available at this website \cite{GitHub}).
We have considered on one side   networks generated from ensembles of Poisson random networks, and on the other side three real network datasets: two  food-webs (Little Rock Lake Food-Web network \cite{Littlerock,data} and Ythan Estuary Food-Web Network \cite{data}) and  the airport network between the top 500 US airports  \cite{USairports}. 
In Figure $\ref{fig:MP_simulations}$ the results of the predictions obtained with the message passing algorithm are compared with simulations of $5,000$ pairs of random realizations of the initial damage for $p^{(1)}_i=p^{(2)}_i=p$ and $p_i^{[11]}=p^2$. 
These results reveal an interesting pattern of the probability $\hat{S}_{[10]}$ that display a clear maximum as a function of $p$. Therefore there is a value of $p$ in which the networks are more unpredictable since the fraction of nodes $\hat{S}_{[10]}$ in the giant component for one realization of the initial  damage but not for the other has a maximum. 
In Figure $\ref{fig:MP_simulations2}$ we display the fraction of nodes $S_{[10]}$ found in the giant component only in one realization of the initial damage in the case of positively correlated, negatively correlated and uncorrelated realizations of the  initial damage.
Two realizations of the initial damage are positively correlated if 
\bea
p^{[11]}_i>p^{(1)}_ip^{(2)}_i,
\eea
for every $i=1,2\ldots, N$. This relation implies that the conditional probability that any given  node is not damaged in the second realization of the initial  damage given that it is not damaged in the first realization, is higher than its unconditional probability.
Similarly two realizations of the initial damage are negatively correlated when 
\bea
p^{[11]}_i<p^{(1)}_ip^{(2)}_i,
\eea
for every $i=1,2\ldots, N$, implying that the conditional probability that any given  node is not damaged in the second realization of the initial  damage given that it is not damaged in the first realization, is smaller than its unconditional probability.

Specifically here  we have considered a damage determined by the following node-independent probabilities, 
\bea
p^{(1)}_i&=&p^{(2)}_i=p, \nonumber \\
p^{[11]}_i&=&p^{a}.
\eea
Here $a\geq 1$ is a parameter tuning the nature of the correlations and such that for $a\in [1,2)$ the two realizations of the damage are positively correlated, for $a>2$ they are negatively correlated and for $a=2$ they are uncorrelated.
Note that while for  $a\le 2$ the range of variability of $p$ is $[0,1]$, for $a>2$ the normalization condition given by Eq. $(\ref{norm})$ limits the largest possible value of $p$ to a number smaller than one.
The results of Figure $\ref{fig:MP_simulations2}$ show that the correlations between two realizations of the initial damage affect the functional relation between the probability $\hat{S}_{[10]}$ and $p$. Notably as a function of the parameter $a$ the maximum of  $\hat{S}_{[10]}$ change position indicating a different value of $p$ in which the system is maximally unpredictable.

Finally from both Figure $\ref{fig:MP_simulations}$ and Figure $\ref{fig:MP_simulations2}$ it is apparent that the message-passing algorithm provides a very good prediction of fluctuations observed in the percolation properties of complex networks. The small deviations observed for some datasets should be attributed to deviations from the locally tree-like assumption.

\section{Fluctuations in random network ensembles} 
\subsection{General equations}
\begin{figure*}[t]
    \includegraphics[width=1.95\columnwidth]{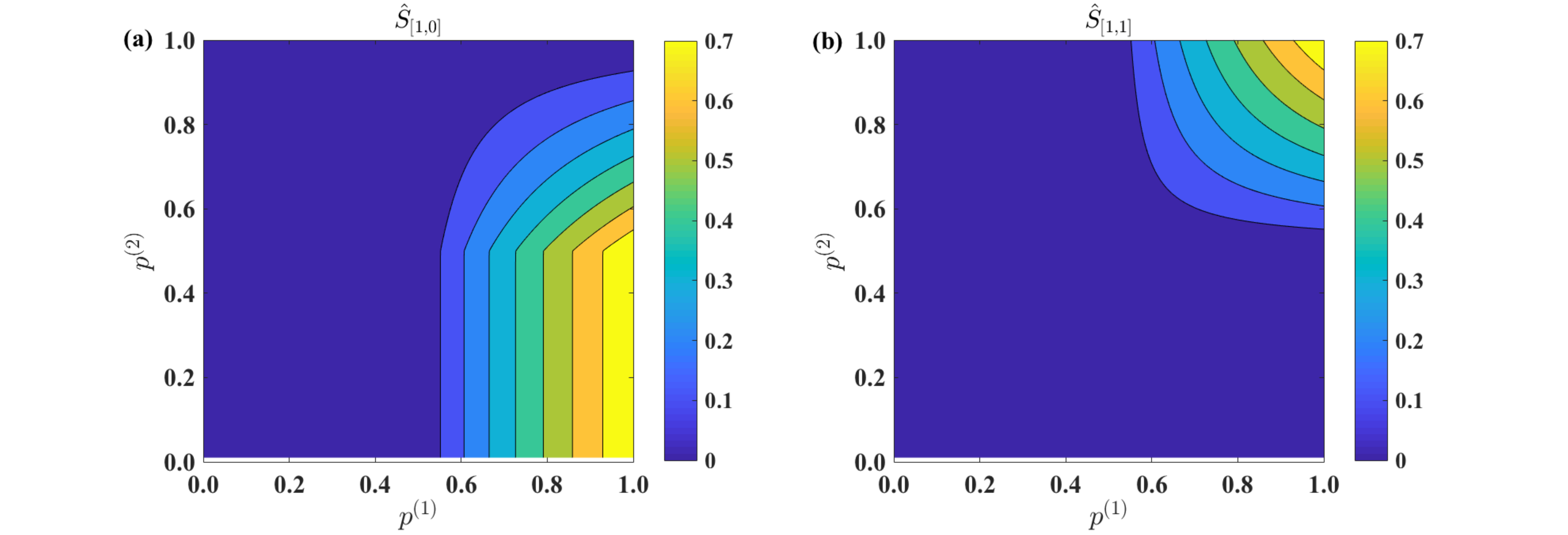}
	\caption{(Color online) The probabilities $\hat{S}_{[10]}$ (panel a) and  $\hat{S}_{[11]}$  (panel b) are plotted as a function of $p^{(1)}$ and $p^{(2)}$ for a Poisson network with average degree $\avg{k}=2$ for the case of two  uncorrelated realizations of the initial damage, i.e. $p^{[11]}=p^{(1)}p^{(2)}$.}
	\label{fig:simple_poisson}
\end{figure*}

On a random uncorrelated network with degree distribution $P(k)$, it is possible not only to predict the expected fraction of nodes $S_{(q)}$ in a given random realization of the initial damage, but is also possible to predict the expected fluctuations by evaluating the expected number of nodes $\hat{S}_{[rr']}$ that are in the giant component in two random realizations of the initial damage (for $[rr']=[11]$) or just in one of the two realizations (for $[rr']=[10]$ and $[rr']=[01]$). This can be achieved  by performing the average of the messages and the indicator functions described in the previous paragraph   over a random uncorrelated network ensemble with given degree distribution $P(k)$ (indicated as $\overline{\ldots}$). To simplify the scenario we consider here and in the following  a pair of realizations of the initial damage where every node $i=1,2,\ldots, N$ is damaged with the same probability, i.e. $p_i^{(q)}=p^{(q)}$ and $p_i^{[11]}=p^{[11]}$.
Therefore,  on locally tree-like uncorrelated network ensembles,  we obtain that $S_{(q)}=\overline{\sigma_i^{(q)}}$, $\hat{S}_{[11]}=\overline{\hat{\sigma}_{i}^{[11]}}$, $\hat{S}_{[10]}=\overline{\hat{\sigma}_{i}^{[10]}}$ and $\hat{S}_{[01]}=\overline{\hat{\sigma}_{i}^{[01]}}$ depend on the values of the  average messages 
$S'_{(q)}=\overline{\hat{\sigma}_{i\to j}^{(q)}}$, $\hat{S}'_{[11]}=\overline{\hat{\sigma}_{i\to j}^{[11]}}$ as indicated by the following equations (see derivation in the Appendix \ref{apA})
	\bea
	S'_{(q)}&=&p^{(q)}\left[1-G_1\left(1-S'_{(q)}\right)\right]\nonumber \\
	S_{(q)}&=&p^{(q)}\left[1-G_0\left(1-S'_{(q)}\right)\right],\nonumber \\
	\hat{S}'_{[11]}&=&p^{[11]}\left[1-G_1\left(1-S'_{(1)}\right)-G_1\left(1-S'_{(2)}\right)\right.\nonumber \\
	&&\left.+G_1\left(1-S'_{(1)}-S'_{(2)}+\hat{S}'_{11}\right)\right],\nonumber \\
	\hat{S}_{[11]}&=&p^{[11]}\left[1-G_0\left(1-S'_{(1)}\right)-G_0\left(1-S'_{(2)}\right)\right.\nonumber \\&&\left.+G_0\left(1-S'_{(1)}-S'_{(2)}+\hat{S}'_{11}\right)\right],\nonumber \\	\hat{S}_{[10]}&=&S_{(1)}-\hat{S}_{[11]},\nonumber \\
	\hat{S}_{[01]}&=&S_{(2)}-\hat{S}_{[11]}
	\label{S12}
	\eea
	with $G_1(z)$ and $G_0(z)$ indicating the generating functions
	\bea
	\begin{array}{lr}
	G_0(z)=\sum_k P(k) z^{k},& G_1(z)=\sum_k\frac{k}{\Avg{k}}P(k) z^{k-1}.
	\label{G01}
	\end{array}
	\eea
	In Figure \ref{fig:simple_poisson} we show the probabilities $\hat{S}_{[10]}$ and $\hat{S}_{[11]}$ as a function of $p^{(1)}$ and $p^{(2)}$ for a Poisson network with average degree $\Avg{k}=2$ and $p^{[11]}=p^{(1)}p^{(2)}$ as predicted by Eqs. $(\ref{S12})$. These plots reveal the entire full-diagram characterizing  the response of the network to external damage.
\\
	\subsection{Two realizations of the initial damage with $p^{(1)}=p^{(2)}=p$}
	In the interesting case in which  the two random realizations of the initial damage have the same probability, i.e. $p^{(1)}=p^{(2)}=p$, the Eqs. $(\ref{S12})$ do simplify significantly  as we have 
	$S'_{(1)}=S'_{(2)}=S'$ and $S_{(1)}=S_{(2)}=S$ . Therefore they reduce to 
		\bea
S'&=&p\left[1-G_1\left(1-S'\right)\right]\nonumber \\
S&=&p\left[1-G_0\left(1-S'\right)\right]\nonumber \\
	\hat{S}'_{[11]}&=&p^{[11]}\left[1-2 G_1\left(1-S'\right)+G_1\left(1-2S'+\hat{S}'_{[11]}\right)\right]\nonumber \\	
	\hat{S}_{[11]}&=&p^{[11]}\left[1-2G_0\left(1-S'\right).+G_0\left(1-2S'+\hat{S}'_{[11]}\right)\right],\nonumber \\
	\hat{S}_{[10]}&=&\hat{S}_{[01]}=S-\hat{S}_{[11]}.
	\label{Sp12}
	\eea
	 
In this case we  observe that both $\hat{S}_{[11]}$ and $\hat{S}_{[10]}$ have a second order phase transition at 	$p=p_c=\frac{\Avg{k}}{\Avg{k(k-1)}}$ where here $\Avg{\ldots}$ indicates the average over the degree distribution $P(k)$ of the network. Let us now characterize the critical behavior of both probabilities on complex networks. These results extend the analysis of the critical indices for percolation of scale-free networks \cite{crit}.
For well behaved distributions with converging first, second and third moment of the degree distribution, as $p\to p_c^{+}$ we  observe the critical behavior
	\bea
S&\propto &\left(p-p_c\right)^{\beta},\nonumber \\
\hat{S}_{[rr']}&\propto&\left(p-p_c\right)^{\hat{\beta}_{[rr']}},
\label{scaling}
\eea
with 
\bea
\begin{array}{cc}
\hat{\beta}_{[11]}=\beta+1,&\hat{\beta}_{[10]}=\beta.
\end{array}
\label{critical2}
\eea
for $p^{[11]}<p$ (which includes the uncorrelated case $p^{[11]}=p^2$),
and 
\bea
\begin{array}{cc}
\hat{\beta}_{[11]}=\beta,&\hat{\beta}_{[10]}=\beta.
\end{array}
\label{critical2b}
\eea
for $p^{[11]}=p$.

Given the fact that for these distributions $\beta$ takes its  mean-field value $\beta=1$ we obtain  $\hat{\beta}_{[11]}=2, \hat{\beta}_{[10]}=1$ for $p^{[11]}<p$ and $\hat{\beta}_{[11]}=1, \hat{\beta}_{[10]}=1$ for $p^{[11]}=p$.
In the relevant case of network with   power-law degree distribution
$	P(k)=C k^{-\gamma}$ and $\gamma>2$  the critical exponents can change and depend on the value of $\gamma$ (see Appendix \ref{apB} for details of the derivation). For $\gamma>4$  we recover the previously discussed scenario as first, second, and third moment of the degree distribution converge. For $\gamma \in(3,4)$ we observe  the scaling of Eq. $(\ref{scaling})$  with critical exponents satisfying Eq. $(\ref{critical2})$ or Eq. $(\ref{critical2b})$ with 
$p_c=\frac{\Avg{k}}{\Avg{k(k-1)}}$, $\beta=\frac{1}{\gamma-3}$. For $\gamma\in (2,3)$ we observe the scaling of Eq. $(\ref{scaling})$ with 
$p_c=0$,  and
\bea
\begin{array}{cc}
\hat{\beta}_{[11]}=\beta+(a-1),&\hat{\beta}_{[10]}=\beta.
\end{array}
\label{critical3}
\eea
for $p^{[11]}=p^a$ and $\beta=\frac{1}{3-\gamma}$.
For $\gamma=4$ we observe logarithmic corrections to the critical behavior
\bea
S&\propto &\left(p-p_c\right)^{{\beta}}\left[-\ln\left(p-p_c\right)\right]^{-1}\nonumber \\
		\hat{S}_{[rr']}&\propto &\left(p-p_c\right)^{\hat{\beta}_{[rr']}}\left[-\ln\left(p-p_c\right)\right]^{-1} \nonumber \\
	\eea
		with $\hat{\beta}_{[11]}$, $\hat{\beta}_{[10]}$ given by Eqs. $(\ref{critical2})$ and $(\ref{critical2b})$ with $\beta=1$ and $p_c=\frac{\Avg{k}}{\Avg{k(k-1)}}$.	
	Finally for  $\gamma=3$ we obtain
	\bea
	S&\propto & p^{\beta} e^{-\frac{c}{p}}\nonumber \\
	\hat{S}_{[rr']}&\propto& p^{\hat{\beta}_{[rr']}}e^{-\frac{c}{p}}\nonumber \\
	\eea
with $c=\avg{k}/C>0$ and the  critical exponents  $\hat{\beta}_{[11]}$,and  $\hat{\beta}_{[10]}$ given by Eqs. $(\ref{critical3})$  with $\beta=1$.\\
	
\section{Conclusions} In conclusion we have presented a characterization of the fluctuations expected in the percolation properties of complex networks. By considering two random realizations of the initial damage, in general correlated,  we  are able to characterize how different nodes might be more stable  than other nodes. Assuming that nodes are damaged randomly with the same probability $f=1-p$ in both realizations of the initial damage, for every single locally tree-like network we have shown how to  predict for which value of $p$  the fluctuations  are more significant both in the case of uncorrelated and correlated realizations of the initial damage. Finally we have studied the percolation on uncorrelated network ensembles characterizing their expected fluctuations. This framework based on a  message-passing algorithm can be applied to single locally tree-like real networks, and here we have discussed its application to food-webs and infrastructure networks. We believe that this approach can be  fruitfully extended to link percolation and to other generalized percolation transitions such as k-core percolation and percolation of  multilayer networks to reveal the role of fluctuations in the response of a network to external damage, also in these generalized scenarios.

\subsubsection*{Note}
Recently we became aware of Ref. \cite{Kuhen} which tackles a  similar problem taking a different perspective.

\appendix
\section{Derivation of Eqs. $(\ref{S12})$ }
\label{apA}

Let us derive here the Eqs. $(\ref{S12})$  for $\hat{S}_{[11]}^{\prime}$ and $\hat{S}_{[11]}^{\prime}$ starting from the message passing Eqs. $(\ref{MP_eq2})$.  A similar    approach can be used to derive the equations for $S^{\prime}_{(q)},S_{(q)}$.
We consider a random realization of the network $G$ drawn from an uncorrelated network ensemble with given degree sequence $\{k_1,k_2,\ldots, k_N\}$, associated to the degree distribution 
\bea
P(k)=\frac{1}{N}\sum_{i=1}^N \delta(k,k_i),
\eea
where $\delta(x,y)$ is the Kronecker delta.
Therefore the network $G$ is  choosen with probability
\bea
P(G)=\frac{1}{Z}\prod_{i=1}^N \delta\left(k_i,\sum_{j=1}^NA_{ij}\right),
\eea
where ${\bf A}$ is its adjacency matrix.

Our aim is to  write the equations for 
the average message $\hat{S}_{[11]}^{\prime}$ and the average probability $\hat{S}_{[11]}$ that a node is in the giant component in both realizations of the percolation problem, i.e.
\bea
\hat{S}^{\prime}_{[11]}&=&\overline{\hat{\sigma}_{i\to j}^{[11]}},\nonumber \\
\hat{S}_{[11]}&=&\overline{\hat{\sigma}_{i}^{[11]}},
\eea
where here we indicated with $\overline{\ldots}$ the average over the probability $P(G)$.
Specifically, for any link-dependent function $f_{i\to j}$ the average $\overline{\ldots}$ indicates 
\bea
\overline{f_{i\to j}}=\sum_{{G}}P({G})\sum_{<i,j>}\frac{f_{i\to j}}{\avg{k}N},
\eea 
where $<i,j>$ are  nearest neighbors.
For node-dependent functions $f_i$, instead $\overline{\ldots}$  indicates the average
\bea
\overline{f_{i}}=\sum_{{G}}P({G})\sum_{i=1}^N\frac{f_{i}}{N}.
\eea 
Using the above definitions together with Eqs. $(\ref{MP_eq2})$ and  the  assumption that the network is locally tree-like, we obtain for $\hat{S}^{\prime}_{[11]}$ 
\bea
\hat{S}^{\prime}_{[11]}&=&\overline{\hat{\sigma}_{i\to j}^{[11]}}\nonumber \\
&=&p^{[11]}\frac{1}{\avg{k}N}\sum_{<i,j>}\left[1- \prod_{\ell \in N(i)\setminus j}\overline{\left(1-  {\sigma}_{\ell \to i}^{(1)} \right)}\right. \nonumber \\
&& -\prod_{\ell \in N(i)\setminus j}\overline{\left(1-  {\sigma}_{\ell \to i}^{(2)} \right)}\nonumber \\
&&+\left.\prod_{\ell \in N(i)\setminus j}\overline{\left(1-  {\sigma}_{\ell \to i}^{(1)}- {\sigma}_{\ell \to i}^{(2)} +\hat{\sigma}_{\ell \to i}^{[11]}\right)}\right]\nonumber\\
&=&p^{[11]}\frac{1}{\avg{k}}\sum_{k}k P(k)\left[1-\left(1-S'_{(1)}\right)^{k-1}\right.\nonumber \\
	&&-\left(1-S'_{(2)}\right)^{k-1}\nonumber \\
	&&+\left.\left(1-S'_{(1)}-S'_{(2)}+\hat{S}'_{11}\right)^{k-1}\right].
 \eea
This equation can be also written as 
\bea
	\hat{S}'_{[11]}&=&p^{[11]}\left[1-G_1\left(1-S'_{(1)}\right)-G_1\left(1-S'_{(2)}\right)\right.\nonumber \\
	&&\left.+G_1\left(1-S'_{(1)}-S'_{(2)}+\hat{S}'_{11}\right)\right],
	\eea
where the generating function $G_1(x)$ is defined in Eq. (\ref{G01}), recovering the Eq. $(\ref{S12})$ for $\hat{S}^{\prime}_{[11]}$.
Similarly, using Eqs. $(\ref{MP_eq2})$ and  the locally tree-like assumption we can calculate $\hat{S}_{[11]}$ getting
\bea
\hat{S}_{[11]}&=&\overline{\hat{\sigma}_{i}^{[11]}}\nonumber \\
&=&p^{[11]}\frac{1}{N}\sum_{i=1}^N\left[1- \prod_{\ell \in N(i)}\overline{\left(1-  {\sigma}_{\ell \to i}^{(1)} \right)}\right. \nonumber \\
&& -\prod_{\ell \in N(i)}\overline{\left(1-  {\sigma}_{\ell \to i}^{(2)} \right)}\nonumber \\
&&+\left.\prod_{\ell \in N(i)}\overline{\left(1-  {\sigma}_{\ell \to i}^{(1)}- {\sigma}_{\ell \to i}^{(2)} +\hat{\sigma}_{\ell \to i}^{[11]}\right)}\right]\nonumber\\
&=&p^{[11]}\sum_{k} P(k)\left[1-\left(1-S'_{(1)}\right)^k-\left(1-S'_{(2)}\right)^k\right.\nonumber \\
	&&\left.+\left(1-S'_{(1)}-S'_{(2)}+\hat{S}'_{11}\right)^k\right].
\eea

This equation can be written in terms of the generating function  $G_0(x)$  defined in Eq. (\ref{G01}) as 
\bea
\hat{S}_{[11]}&=&p^{[11]}\left[1-G_0\left(1-S'_{(1)}\right)-G_0\left(1-S'_{(2)}\right)\right.\nonumber \\&&\left.+G_0\left(1-S'_{(1)}-S'_{(2)}+\hat{S}'_{11}\right)\right],
\eea
recovering the Eq. $(\ref{S12})$ for $\hat{S}_{[11]}$.

\section{Derivation of the critical indices}
\label{apB}
In this appendix we give the details of the derivation of the critical indices.
\subsection*{Well behaved degree distributions}
In this paragraph we derive the critical indices in  the case of well behaved degree distributions $P(k)$ having first, second and third convergent moment.
Starting from Eqs. $(\ref{Sp12})$, and expanding close to the trivial solution
$S=S^{\prime}=\hat{S}_{[r,r']}=0$ we get
\bea
S^{\prime}&=&p\frac{\Avg{k(k-1)}}{\avg{k}}S^{\prime}-p\frac{1}{2}\frac{\Avg{k(k-1)(k-2)}}{\avg{k}}(S^{\prime})^2+\ldots\nonumber \\
S&=&p{\avg{k}}S^{\prime}+\ldots\nonumber \\
\hat{S}_{[11]}^{\prime}&=&p^{[11]}\frac{\Avg{k(k-1)}}{\avg{k}}\hat{S}_{[11]}^{\prime}\nonumber \\
&&+p^{[11]}\frac{1}{2}\frac{\Avg{k(k-1)(k-2)}}{\avg{k}}\nonumber \\
&&\times\left[-2(S^{\prime})^2+\left(-2S^{\prime}+\hat{S}_{[11]}^{\prime}\right)^2\right]+\ldots\nonumber \\
\hat{S}_{[11]}&=&p^{[11]}\avg{k}\hat{S}_{[11]}^{\prime}+\ldots\nonumber 
\eea
Considering the first relevant terms of the expansion, we find for $S$ and $S^{\prime}$
\bea
S^{\prime}&\propto&\left(p\frac{\Avg{k(k-1)}}{\avg{k}}-1\right),\nonumber \\
S&\propto&\left(p\frac{\Avg{k(k-1)}}{\avg{k}}-1\right),
\eea
as long as $S\ll 1$ and $S^{\prime}\ll 1$.
When investigating for  the scaling for $\hat{S}_{[11]}^{\prime}, \hat{S}_{[11]}$ we need to distinguish between the cases: $p^{[11]}<p$ and $p^{[11]}=p$.
In the case  $p^{[11]}<p$ we have for $\hat{S}_{[11]}\ll 1$ and $\hat{S}^{\prime}_{[11]}\ll 1$
\bea
\hat{S}^{\prime}_{[11]}&\propto&\left(p\frac{\Avg{k(k-1)}}{\avg{k}}-1\right)^2\nonumber \\
\hat{S}_{[11]}&\propto&\left(p\frac{\Avg{k(k-1)}}{\avg{k}}-1\right)^2\nonumber \\
\eea
In the case $p^{[11]}=p$ we have instead for $\hat{S}_{[11]}\ll 1$ and $\hat{S}^{\prime}_{11}\ll 1$
\bea
\hat{S}^{\prime}_{[11]}&\propto&\left(p\frac{\Avg{k(k-1)}}{\avg{k}}-1\right)\nonumber \\
\hat{S}_{[11]}&\propto&\left(p\frac{\Avg{k(k-1)}}{\avg{k}}-1\right)\nonumber \\
\eea
Therefore $S$,$S_{[10]}=S-S_{[11]}$ and $S_{[11]}$, close to the transition point ($p\to p_c^{+}$), follow the scaling 
\bea
S&\propto &\left(p-p_c\right)^{\beta},\nonumber \\
\hat{S}_{[rr']}&\propto&\left(p-p_c\right)^{\hat{\beta}_{[rr']}},
\label{scalinga}
\eea
with $p_c=\frac{\avg{k}}{\avg{k(k-1)}}$, $\beta=1$ and
\bea
\begin{array}{cc}
\hat{\beta}_{[11]}=\beta+1,&\hat{\beta}_{[10]}=\beta.
\end{array}
\label{critical2a}
\eea
in the case $p^{[11]}<p$, and  
\bea
\begin{array}{cc}
\hat{\beta}_{[11]}=\beta,&\hat{\beta}_{[10]}=\beta.
\end{array}
\label{critical2ba}
\eea
in the case $p^{[11]}=p$.
\subsection*{Power-law degree distributions}
In this paragraph we derive the critical indices in  the case of a power-law degree distribution $P(k)=Ck^{-\gamma}$ where  $\gamma>2$ and $C$ is the normalization constant.
\subsubsection*{Case $\gamma>4$}
In the case in which $\gamma>4$ the degree distribution $P(k)$ has converging first, second and third  moment. Therefore this case can be recast in the case of well behaved distributions discussed above.

\subsubsection*{Case $\gamma=4$}
 Expanding Eqs. $(\ref{Sp12})$ close to the trivial solution
$S=S^{\prime}=\hat{S}_{[r,r']}=0$ we obtain, for $\gamma=4$,
\bea
S^{\prime}&=&p\frac{\Avg{k(k-1)}}{\avg{k}}S^{\prime}+pD (S^{\prime})^{2}\ln S^{\prime}+\ldots,\nonumber \\
S&=&p{\avg{k}}S^{\prime}+\ldots,\nonumber \\
\hat{S}_{[11]}^{\prime}&=&p^{[11]}\frac{\Avg{k(k-1)}}{\avg{k}}\hat{S}_{[11]}^{\prime}+\ldots, \nonumber \\
&&+p^{[11]} D\left[2(S^{\prime})^{2}\ln S^{\prime}-(2S^{\prime}-\hat{S}_{[11]}^{\prime})^{2}\ln (2S^{\prime}-\hat{S}_{[11]}^{\prime})\right]+\ldots,\nonumber \\
\hat{S}_{[11]}&=&p^{[11]}\avg{k}\hat{S}_{[11]}^{\prime}+\ldots,
\eea
where $D>0$ is a constant.
Proceeding as in the precent case we recover, close to the transition point, the scaling behavior
\bea
S&\propto&\left(p-p_c\right)^{{\beta}}\left[-\ln\left(p-p_c\right)\right]^{-1}
\nonumber \\
		\hat{S}_{[rr']}&\propto &\left(p-p_c\right)^{\hat{\beta}_{[rr']}}\left[-\ln\left(p-p_c\right)\right]^{-1} \nonumber \\
	\eea
		with
		\bea
		\beta &=&1,\nonumber \\ 
		p_c&=&\frac{\Avg{k}}{\Avg{k(k-1)}},
		\eea
		and  $\hat{\beta}_{[11]}$, $\hat{\beta}_{[10]}$ given by Eqs. $(\ref{critical2a})$ and $(\ref{critical2ba})$.
		
\subsubsection*{Case $\gamma\in (3,4)$}
In the case $\gamma\in (3,4)$, starting from Eqs. $(\ref{Sp12})$, and expanding close to the trivial solution
$S=S^{\prime}=\hat{S}_{[r,r']}=0$ we get
\bea
S^{\prime}&=&p\frac{\Avg{k(k-1)}}{\avg{k}}S^{\prime}-pD (S^{\prime})^{\gamma-2}+\ldots,\nonumber \\
S&=&p{\avg{k}}S^{\prime}+\ldots,\nonumber \\
\hat{S}_{[11]}^{\prime}&=&p^{[11]}\frac{\Avg{k(k-1)}}{\avg{k}}\hat{S}_{[11]}^{\prime}+\ldots, \nonumber \\
&&+p^{[11]}D\left[-2(S^{\prime})^{\gamma-2}+(2S^{\prime}-\hat{S}_{[11]}^{\prime})^{\gamma-2}\right]+\ldots,\nonumber \\
\hat{S}_{[11]}&=&p^{[11]}\avg{k}\hat{S}_{[11]}^{\prime}+\ldots,
\eea
where $D>0$ indicates a constant.
Close to the transition point, we recover the scaling behavior Eq. $(\ref{scalinga})$ and the critical indices determined by Eqs. $(\ref{critical2a}),(\ref{critical2ba})$ with 
\bea
\beta&=&\frac{1}{\gamma-3},\nonumber \\
p_c&=&\frac{\avg{k}}{\Avg{k(k-1)}}.
\eea

\subsubsection*{Case $\gamma=3$}
Expanding Eqs. $(\ref{Sp12})$ close to the trivial solution
$S=S^{\prime}=\hat{S}_{[r,r']}=0$ for $\gamma=3$ we obtain
\bea
S^{\prime}&=&-p \frac{C}{\avg{k}} (S^{\prime})\ln (S^{\prime})+\ldots,\nonumber \\
S&=&p{\avg{k}}S^{\prime}+\ldots,\nonumber \\
S_{[11]}^{\prime}&=&p^{[11]} \frac{C}{\avg{k}}\left[2(S^{\prime})\ln(S^{\prime})\right.\nonumber \\
&&\left.-(2S^{\prime}-S_{[11]}^{\prime})\ln\left(2S^{\prime}-S_{[11]}^{\prime}\right)\right]+\ldots,\nonumber \\
S_{[11]}&=&p^{[11]}\avg{k}S_{[11]}^{\prime}+\ldots, \nonumber \\
\eea
These expressions yield the scaling
\bea
	S&\propto & p^{\beta} e^{-\frac{c}{p}}\nonumber \\
	\hat{S}_{[rr']}&\propto& p^{\hat{\beta}_{[rr']}}e^{-\frac{c}{p}}\nonumber \\
	\eea
	with $c=\avg{k}/C>0$, $\beta=1$
	and critical indices 
\bea
\beta_{[10]}&=&\beta,\nonumber \\
\beta_{[11]}&=&\beta+(a-1),
\eea
for $p^{[11]}=p^a$ with $a\geq 1$.

\subsubsection*{Case $\gamma\in (2,3)$}
Expanding Eqs. $(\ref{Sp12})$ close to the trivial solution
$S=S^{\prime}=\hat{S}_{[r,r']}=0$ we obtain
\bea
S^{\prime}&=&p {D} (S^{\prime})^{\gamma-2}+\ldots,\nonumber \\
S&=&p{\avg{k}}S^{\prime}+\ldots,\nonumber \\
S_{[11]}^{\prime}&=&p^{[11]} {D}\left[2(S^{\prime})^{\gamma-2}-(2S^{\prime}-S_{[11]}^{\prime})^{\gamma-2}\right]+\ldots,\nonumber \\
S_{[11]}&=&p^{[11]}\avg{k}S_{[11]}^{\prime}+\ldots, \nonumber \\
\eea
where ${D}>0$ indicates a constant.
This expression yield the scaling behavior defined in Eq.$(\ref{scalinga})$ with 
\bea
\beta &= & \frac{1}{3-\gamma}\nonumber \\
p_c&=&0
\eea
and critical indices 
\bea
\beta_{[10]}&=&\beta,\nonumber \\
\beta_{[11]}&=&\beta+(a-1),
\eea
for $p^{[11]}=p^a$ with $a\geq 1$.

\end{document}